\documentclass[11pt]{article}

\usepackage{amsmath, amsfonts, amssymb, bm, bbm,color, amsthm, enumerate}

\usepackage{amsmath, amsfonts, amssymb, bm, bbm,color, amsthm, fullpage}

\usepackage{multirow}
\usepackage[authoryear, round]{natbib}
\usepackage{hyperref}

\usepackage{tikz}
\newcommand{\td}{\mathrm{d}}

\newcommand{\pr}{\mathbb{P}}
\newcommand{\ep}{\mathbb{E}}

\usepackage{setspace}

\newcommand{\lp}{{\cal L}}

\newcommand{\im}{\mathrm{Im}}

\newcommand{\X}{{\cal X}}
\newcommand{\Y}{{\cal Y}}

\newcommand{\I}{\mathbbm{i}}

\usepackage{helvet}
 
\usepackage[T1]{fontenc}

\newtheorem{theorem}{Theorem}

\newtheorem{definition}{Definition}

\begin{document}

\title{Unseeded low-rank graph matching by transform-based unsupervised point registration}
\author{Yuan Zhang\\\\Department of Statistics,  The Ohio State University\\229 Cockins Hall, 1958 Neil Avenue, Columbus, Ohio, USA\\\\yzhanghf@stat.osu.edu}

\maketitle

\begin{abstract}
	The problem of learning a correspondence relationship between nodes of two networks has drawn much attention of the computer science community and recently that of statisticians.  The unseeded version of this problem, in which we do not know any part of the true correspondence, is a long-standing challenge.  For low-rank networks, the problem can be translated into an unsupervised point registration problem, in which two point sets generated from the same distribution are matchable by an unknown orthonormal transformation.  Conventional methods generally lack consistency guarantee and are usually computationally costly.
	
	In this paper, we propose a novel approach to this problem.  Instead of simultaneously estimating the unknown correspondence and orthonormal transformation to match up the two point sets, we match their distributions via minimizing our designed loss function capturing the discrepancy between their Laplace transforms, thus avoiding the optimization over all possible correspondences.  This dramatically reduces the dimension of the optimization problem from $\Omega(n^2)$ parameters to $O(d^2)$ parameters, where $d$ is the fixed rank, and enables convenient theoretical analysis.  In this paper, we provide arguably the first consistency guarantee and explicit error rate for general low-rank models.  Our method provides control over the computational complexity ranging from $\omega(n)$ (any growth rate faster than $n$) to $O(n^2)$ while pertaining consistency.  We demonstrate the effectiveness of our method through several numerical examples.
\end{abstract}

\section{Introduction}

The problem of estimating the correspondence between two graphs has a long history and has a wide range of applications, including multiple-layer social network analysis, pattern recognition and computer vision, biomedical image analysis, document processing and analysis and so on.  For a comprehensive review of these and more applications, see \citet{conte2004thirty} and \citet{fishkind2012seeded}.

The prototype graph matching problem has the following basic form.  Suppose we have participating individuals numbered as $1,\ldots,n$ in the network (or called ``graph'' -- in this paper, we shall use ``network'' and ``graph'' interchangeably) as nodes or vertices.  The data we collect describe the interactions or relationships between them, called edges.  The edges may be binary or weighted, depending on the context, and the same set of individuals form two networks, but at least one of the networks has the order of its nodes shuffled, and the correspondence between the nodes of the two networks is missing.  The primary goal of graph matching problem is to recover the lost node mapping, such that after aligning the order of nodes in one network to those in the other network would result the same or a similar networks.  The data we observe are the two networks with their node orders shuffled, and further the data may be contaminated by random noises, as we shall explain right next.

There are two main versions of this problem.  The exact graph matching problem assumes no randomness or noise in graph generation but only that the two graphs are exactly identical under the hidden true node correspondence.  The task is to recover the true map.  It is well-known to be an NP problem, despite the recent significant advances \citep{babai2016graph, svensson2017matching} showing that it could be solved in quasi-polynomial time.  The other version is inexact graph matching.  This version assumes that data are observed with random noise.  For example, a popular assumption is that the true graphs are edge probability matrices and only their Bernoulli realizations are observable -- moreover, the generations of the corresponding edges in the two graphs may be dependent, such as the model studied in \citet{lyzinski2014seeded}.

The existing research on the exact and inexact graph matching problems is not superficially nested as the appearance of them may suggest, but rather pointing to distinct directions.  Research on the former largely focused on the worst-case complexity; while the latter has usually been discussed with structural assumptions.  Network data with $n$ nodes involved apparently has $\asymp n^2$ complexity.  But with structures assumed, this is significantly reduced.  In this paper, we shall concentrate our attentions on the low-rank case, in which one may roughly think that data essentially reside in $O(nd)$ space, where $d$ is the dimension of assumed structures, thus solving the problem efficiently is possible.  The low-rank model is more universal than it seems.  Letting the rank grow, we may hope to consistently approach very general network structures \citep{bickel2009nonparametric, gao2015rate, xu2017rates}.  Recent advancements in matrix analysis further suggest that low-rank models are decent approximations to much more general models, for example, \citet{udell2017nice} claims that smooth structures can be approximated by models with $\log(n)$ rank growth rate.

By far, we have been discussing the unsupervised graph matching problem.  Its counterpart -- the seeded graph matching problem is also a popular topic.  Seeds refer to subsets of nodes in the two networks, the true correspondence relationship between the members of which are known.  It is intuitively understandable that a ``representative'' pair of seed node set, even with small cardinality, may dramatically lower the difficulty of the problem, making it not only polynomial.  It is known that the seeded graph matching problems for graphs with low-rank structures are usually efficiently solvable \citep{lyzinski2014seeded, lyzinski2014seeded2, lyzinski2014seeded3}.  Learning any seed node in the unseeded context, however, seems difficult, as an efficient method that solves this problem may lead to P=NP.

In existing literature to date, the difficulty of the unseeded graph matching problem remains unknown.  Despite the conjecture that it should be efficiently solvable, currently there exists no such provable method.  The unknown node correspondence has been playing the main obstacle in the way.  The problem can be translated into a point registration problem in $\mathbb{R}^d$ space, but the two point clouds to be matched are further gapped by an unknown orthonormal transformation on one of them, thus it cannot be solved by directly applying a Hungarian algorithm.  Attempts to solve this problem by far almost all involve an optimization partially over the unknown correspondence, and the focuses have been relaxing it.  This makes these methods hard to analyze and many of their computations costly.

In this paper, we present a novel method that solves this problem in polynomial time with theoretical guarantees under quite mild structural assumptions.  Our approach is distinct from the majority, if not all, of existing methods in that we directly pursue the low-rank structure thus completely avoided the optimization over the $n\times n$ permutation matrix in the main stages of our method -- only except that we run the Hungarian algorithm or its alternative \emph{just once} at the end.  Our method is simple, scalable and convenient to analyze, not only utilizing our analysis, but potentially also enabling a rich variety of subsequent estimations and inferences.

\section{Problem formulation}

We represent a graph of $n$ nodes by its $n$ by $n$ adjacency matrix $A$, where $A_{ij}=1$ if there is an edge from node $i$ to node $j$, and $A_{ij}=0$ otherwise.  For simplicity, in this paper we only discuss binary and symmetric graphs with independent edge generations, that is,  for every $i,j$ pair such that $i\leq j$, generate data by $A_{ij}=A_{ji}\sim$~Bernoulli$(W_{ij})$, where $W\in[0,1]^{n\times n}$ is the edge probability matrix, and for every $(i,j)\neq (i',j')$, where $i\leq j$ and $i'\leq j'$, $A_{ij}$ is independent of $A_{i'j'}$.  This is a popular model basis studied in many network inference papers such as \citet{bickel2009nonparametric, wolfe2013nonparametric, gao2015rate} and \citet{zhang2017estimating}.

If the edge probabilities are completely arbitrary numbers and unrelated to each other, no meaningful inferences would be possible, so now we quantitatively define what we mean by ``network structures''.  According to the Aldous-Hoover representation \citep{aldous1981representations, hoover1979relations}, the edge probability matrix of an exchangeable network can be written as:

\begin{definition}[Aldous-Hoover]
	For any exchangeable network, there exists a symmetric function $f:[0,1]^2\to[0,1]; f(x,y)=f(y,x)$, called the ``graphon function'', and a set of i.i.d. random variables $u_1,\ldots,u_n\sim$~Uniform$[0,1]$, such that the edge probability matrix $W$ can be represented by
	$$
	W_{ij} = f(u_i, u_j)
	$$
\end{definition}
For directed co-exchangeable networks, simply remove the symmetry requirement on $f$ and $W_{ij}$ can be represented by $W_{ij}=f(u_i,v_j)$, where $u$ and $v$'s are independent standard uniform random variables.   Notice that both $f$ and the latent positions $u$, and $v$ if applicable, are not-estimable due to identifiability issues, unless some strong additional model assumptions are assumed \citep{airoldi2013stochastic}.  Indeed, many existing work on graphon estimation tend to assume smoothness on $f$, so will this paper, but such assumptions usually only help us in indirect ways, such as elucidated in \citet{gao2015rate} and \citet{zhang2017estimating}.  Notice that the smoothness in $f$ does not mean that the resulting distribution of the elements of $W$ is continuous -- a quick example is the Erd\"os-Renyi model, in which $f(x,y)\equiv p\in(0,1)$.

The Aldous-Hoover representation has a more specific form for low-rank networks.  Here we impose our low-rank assumption on $f$, and the low-rankness is straightforwardly inherited by the probability matrix $W$ generated based on $f$.  We have the functional spectral decomposition of $f$ as follows:
\begin{equation}
f(x,y) = \sum_{k=1}^d \lambda_k \xi_k(x)\xi_k(y)
\label{eqn::graphon::spectral_decomposition}
\end{equation}
where $\lambda_k$ is the $k$th largest nonzero eigenvalue and $\xi_k$ is its corresponding eigenfunction that is defined on $[0,1]$ and $\int_0^1\xi_k^2(t)\td t=1$.  In this paper, we only consider piece-wise Lipschitz $f$ universally bounded between 0 and 1, and one can show that this implies the universal boundedness of all the eigenfunction $\xi_k$'s piece-wise Lipschitz and thus their are universal boundedness.

Now, based on \eqref{eqn::graphon::spectral_decomposition}, we may represent $W$ from a low-rank graphon as follows:
\begin{equation}
W=\Xi \Lambda \Xi^T = XJX^T
\label{eqn::prob_mat::low-rank-graphon}
\end{equation}
where $\Xi, X\in\mathbb{R}^{n\times d}$ and $\Lambda, J\in\mathbb{R}^{d\times d}$ are defined as $\Xi_{i, k}:=\xi_k(u_i)$, $X_{i,k}:=\sqrt{|\lambda_k|}\xi_k(u_i)$, $\Lambda:=\textrm{diag}(\lambda_1,\ldots,\lambda_d)$ and $J:=\textrm{diag}(\textrm{sign}(\lambda_1),\ldots,\textrm{sign}(\lambda_d))$, respectively.  Similar representation also appeared recently in \citet{lei2018network}.  When the graph $W$ is positive semi-definite (PSD) then $J=I$ and simply
$$
W = XX^T
$$
This model is called random dot-product graph (RDPG) \citep{young2007random, athreya2017statistical}.  For general $J=\textrm{blockdiag}(I_{d_1},-I_{d_2})$, where $d_1+d_2=d$, we may separately estimate for the positive- and negative-semidefinite parts.  For simplicity of illustration, in the statement of our method and the theory, we focus on the PSD case for simplicity.

Now we are ready to formally introduce the graph matching problem in the low-rank setting.  Suppose we have two graphs $A^{(1)}$ and $A^{(2)}$ generated based on the same low-rank probability matrix $W=XJX^T$ as in \eqref{eqn::prob_mat::low-rank-graphon}, but the rows and columns of the second network is permuted by an unknown permutation $P^*\in\{0,1\}^{n\times n}, \left(P^*\right)^TP^*=I$.  Denoting the two edge probability matrices by $W^{(1)}$ and $W^{(2)}$, we have
$$
W^{(1)} = W,\quad\textrm{and}\quad W^{(2)} = P^*W^{(1)}\left(P^*\right)^T
$$
and defining $Y:=P^* X$, the induced data generation can be described as follows:
\begin{align}
A^{(1)}_{ij}&=A^{(1)}_{ji}\sim \textrm{Bernoulli}(W_{ij}) = \textrm{Bernoulli}((XJX^T)_{ij})  \label{A1::generate}\\
A^{(2)}_{ij}&=A^{(2)}_{ji} \sim \textrm{Bernoulli}((P^*W(P^*)^T)_{ij})=:\textrm{Bernoulli}((YJY^T)_{ij})  \label{A2::generate}
\end{align}
where $(i,j)$ ranges over all index pairs satisfying $i\leq j$.

If we have access to $W$ and the nonzero eigenvalues are distinct, then we may exactly recover $X$ and $Y$ only up to an $\pm1$ multiplier on each of its columns.  For not-too-large $d$, we can exhaust all $2^d$ possible sign flip combinations on the $d$ columns of $X$, and, for each of them, we run a Hungarian algorithm to match the rows of the column-wise sign flipped $X$ to $Y$.  This further leads to an exact recovery of the correspondence true node correspondence $P^*$.  But the problem would seemingly grow significantly less trivial, even in the oracle, if $W$ has repeated nonzero eigenvalues.  The estimation may only get to the spanning linear space of the corresponding columns in $X$, and now the rows of $X$ and $Y$ are only matchable up to an unknown orthonormal transformation $O^*$ on the columns, that is $P^*XO^* = Y$.  Another source that contribute to the introduction of the latent orthonormal transformation is the concentration inequalities regarding $X$ and $Y$.  In practice, we never observe $W$ and may only work $A$ and the estimated $\hat{X}$ from decomposing $A$.  By Davis-Kahan type theorem \citep{yu2014useful} and concentration results of eigenvalues, we can only approximate $X$ by $\hat{X}$ from $A\approx \hat{X}\hat{X}^T$ up to an unknown transform $\hat{O}_X$ such that $\|X- \hat{X}\hat{O}_X\|_F = O_p(1/\sqrt{n})$.

Now it is clear that the unseeded low-rank graph matching problem can be translated into an unsupervised point registration problem.  Suppose there are two sets of points in a bounded set of $\mathbb{R}^d$.  The two data sets $\{ x_1,\ldots,x_n \}$ and $\{y_1,\ldots,y_n\}$ are i.i.d. samples random vectors $\X$ and $\X O^*$, respectively, where $O^*\in\mathbb{R}^d$ is an unknown orthonormal transformation.  In this paper, distinct from most existing work, we do not impose any smoothness assumption on the distribution of $\X$, but instead only assume its universal boundedness, which is naturally satisfied when the point registration problem origins from the low-rank graph matching problem.  The main task is to estimate both the transform $O^*$ and the permutation matrix $P^*$ that minimize the MSE loss function:
\begin{equation}
\min_{P,O}\|PX^{(1)}O-X^{(2)}\|_F
\label{problem::point_registration}
\end{equation}
where the rows of $X^{(1)}$ and $X^{(2)}$ are $x_i$ and $y_j$'s.  As mentioned earlier, we may not have access to $X^{(1)}$ and $X^{(2)}$, but instead only observe error-contaminated versions of them.  Moreover, the measurement errors may be dependent across sample points, but in fact this does not pose additional challenge to our method.  For this reason, when introducing our method, we focus our attentions on the vanilla form of the unsupervised point registration problem \eqref{problem::point_registration}.

\section{Related work}


In this section, we briefly review some popular existing methods for point registration and graph matching, respectively.  Arguably one of the most popular point registration methods is Iterative Closest Point (ICP) \citep{ezra2006icp, du2010affine, maron2016point}.  It solves the optimization problem \eqref{problem::point_registration} by iteratively optimizing over $P$ and $O$.  This method is simple yet popular.  An ICP equipped with Hungarian algorithm costs $O(n^3)$ in each iteration, making it hard for large data sets.  Another popular method is kernel correlation (KC).  KC matches the two distributions by minimizing the integrated difference between their density functions empirically approximated by kernel density estimations (KDE).  KC is originally designed only for continuous distributions, and it has a distinct form for discrete distributions.  It is substantively difficult to apply KC to distributions of mixed continuity types.

Many existing methods on graph matching are based on seed nodes.  Representative seed nodes may significantly reduce the difficulty of the problem and allows for efficient method for estimating the matching of graphs of general structures.  On the other hand, most existing methods for unseeded graph matching focus on relaxing $P$ in the following optimization problem
\begin{equation}
\min_{P}\|PA^{(1)}P^T - A^{(2)}\|_F
\label{problem::graph_matching}
\end{equation}
from permutation into a continuum such as doubly stochastic relaxations, see \citet{lyzinski2016information} and \citet{vogelstein2015fast}.  As suggested by \citet{lyzinski2016graph}, convex relaxations on $P$ almost never find the global optimality unless initialized already close to the optimal solution.

\section{Our method}

To introduce our method, we start with the observation that the main challenge in solving \eqref{problem::point_registration} lies in the optimization over the permutation matrix $P$.  This is a chicken-and-egg problem.  Notice that if either the optimal $O=O^*$ or even part of the optimal $P=P^*$ is known or well-estimated, the estimation of the remaining parameters would be greatly simplified.  This motivates us to consider the possibility of estimating only one of them and bypassing the optimization over the other one.  Between $P$ and $O$, clearly $O$ is a more ``essential'' parameter, because $P$ may look very differently from realization to realization and even have different dimensions if we consider the more general version of the point registration problem with different sample sizes; where as $O$ determines how the two \emph{distributions} should be distorted to match up with each other.

The core idea of our method is that instead of aiming at matching up the individual points, we match the two distributions.  To serve this purpose, we design a discrepancy measure that describes the difference between the two distributions as a function of $O$.  This naturally introduces an optimization problem over only $O$, circumventing the optimization over $P$ since the empirical version of any such discrepancy measure would depend on data only through the empirical distributions of $\X$ and $\Y$, invariant to the order in which we observe the individual points and thus invariant to $P$.

We now focus on the design of the discrepancy measure between distributions.  Recall that we desire this measure to be well-defined for all distribution continuity types.  One natural choice is to match their moments.  Specifically, we want to match \emph{all} their moments simultaneously, since for any $k\in\mathbb{N}^+$, one may always find random vectors $\X_0$ and $\Y_0$ such that all their $1,\ldots,k$th moments match, but at least one of their $k+1$st moments do not match.  This naturally leads us to consider moment-generating transformations.

Among the arguably most popular choices, including moment generating function (MGF), Laplace transform and characteristic function (CF), we choose to work with Laplace transform for its convenient inversion formula form that significantly facilitates theoretical analysis.  MGF's known inversion formula \citep{post1930generalized, widder2015laplace} is an infinite series; while CF's inversion formula for recovering cumulative distribution function (CDF) is defined in a limit form (L\'evy's theorem, see \citet{durrett2010probability}).  The complicated CDF inversion formula of CF brings technical obstacles in analysis.  Conventional Laplace transforms are defined only for positive random variables and vectors, but we will see that both the Laplace transform and its inversion formula are well-defined for universally bounded random variables and vectors, too.  For a random vector $\X \in \mathbb{R}^d$ satisfying $\|\X\|_2\leq M$ for a universal constant $M>0$, its Laplace transform is defined by
\begin{equation}
\lp_\X(s) = \ep\left[ e^{-\langle s,\X \rangle} \right] = \ep\left[ e^{-(s_1\X_1+\cdots+s_d\X_d)} \right]
\label{definition::Laplace_transform}
\end{equation}
where $s\in\mathbb{C}^d$.  The inversion formula for \eqref{definition::Laplace_transform} that recovers $\X$'s joint CDF is
\begin{equation}
F_{\X}(t)=\pr(\X\leq t) = \frac{1}{(2\pi)^d} \lim_{T_1,\ldots,T_d\to\infty} e^{\langle t,s \rangle} \int_{\prod_{k=1}^d\left[ \gamma-\I T_k,\gamma+\I T_k  \right]} \frac{\lp_{\X}(s)}{s_1\cdots s_d}\td s_1\cdots\td s_d
\label{definition::Inverse_Laplace_transform}
\end{equation}
where the integration limit $\left[ \gamma-\I T_k,\gamma+\I T_k  \right]$ means integrating $s_k$ on the line segment connecting the two points $\gamma\pm\I T_k$.  Given two random vectors $\X O$ and $\Y$, where the former is tuned by an orthonormal transform $O\in\mathbb{R}^{d\times d}$, we wish to estimate $O$ that matches these two distributions.  For this purpose, we define a loss function that describes the discrepancy between the two functions $\lp_{\X O}(s)$ and $\lp_\Y(s)$.   Inspired by \eqref{definition::Inverse_Laplace_transform}, we design the population version of our loss function as follows
\begin{equation}
\Delta(O;\X,\Y;R):= \int_{[\gamma-\I R, \gamma+\I R]^d}\frac{\left|\lp_{\X O}(s)-\lp_\Y(s)\right|}{\left| s_1\cdots s_d \right|}\td s_1\cdots\td s_d
\label{definition::loss_func_population}
\end{equation}
where $R>0$ is a tuning parameter that will be set by the theory.  Clearly, under our assumption that the two distributions under study are matchable, the only $O$'s that achieve the minimum of 0 of \eqref{definition::loss_func_population} for all $R>0$ are those that match the distribution of $\X$ to $\Y$, that is, $\X O\stackrel{d}{=}\Y$.  The form \eqref{definition::loss_func_population} is intractable since it contains unknown components $\lp_{\X O}(s)$ and $\lp_\Y(s)$ and their integration over a continuum.  Therefore, in practice, we work with its sample version.  In order to realize the integration over $s$, we sample $s$ by the following importance sampling:
\begin{itemize}
	\item Define
	$$
	\widetilde{C}(R;\gamma):=\log\left[ 1+2R\left\{ R+(R^2+\gamma^2) \right\}/\gamma^2 \right]
	$$
	Clearly
	$$
	\widetilde{C}(R;\gamma) \asymp \log R
	$$
	as $R\to\infty$ with a fixed $\gamma>0$.
	\item $s_{j_s}=(s_{j_s,1},\ldots,s_{j_s,d})$, where $s_{j_s,k}$ is independent of any other $s_{j_s,\ell}$, for all $\ell\neq k$
	\item For each $k$, set $s_{j_s,k} = \gamma+\I t_{j_s,k}$, where $\gamma>0$ is a preset constant, and the imaginary part $t_{j_s,k}$ is sampled from the continuous distribution with the density function $\pi(t)$:
	$$
	\pi(t):=\frac{1}{\widetilde{C}(R;\gamma)\cdot (\gamma^2+t^2)^{1/2}}\mathbbm{1}_{[|t|\leq R]}
	$$
\end{itemize}
The importance sampling scheme reflects the fact that as the imaginary part of $s$ drifts away from 0, the influence of the Laplace transform $\lp_X(s)$ on the shape of the CDF function $F_\X(t)$ decreases.  We are now ready to define the sample version of our loss function \eqref{definition::loss_func_population} as follows:
\begin{equation}
\widehat{\Delta}(O;X,Y;R) := \frac{1}{m_s}\sum_{j_s=1}^{m_s}\left|  \frac{1}{n_X}\sum_{j_X=1}^{n_X} e^{-\langle s_{j_s},X_{j_X\cdot}O \rangle}  -  \frac{1}{n_Y}\sum_{j_Y=1}^{n_Y} e^{-\langle s_{j_s},Y_{j_Y\cdot} \rangle}  \right| \cdot \left\{ \widetilde{C}(R;\gamma) \right\}^d
\label{definition::loss_func_sample}
\end{equation}
where the rows of $X$ and $Y$ are independent samples from the distributions of $\X$ and $\Y$, respectively, that is:
$$
X_{j_X\cdot}\sim{\cal F}_\X\quad \textrm{and}\quad Y_{j_Y\cdot}\sim{\cal F}_\Y
$$
In practice, the factor $\left\{ \widetilde{C}(R;\gamma) \right\}^d$ introduced by importance sampling in \eqref{definition::loss_func_sample} can be ignored.  Notice that $\widehat{\Delta}(O;X,Y;R)$ is smooth for all $O$ that $\widehat{\Delta}(O;X,Y;R)\neq 0$. After $O$ is estimated, we may simply run a Hungarian algorithm to obtain the mapping between the points.  In the unseeded low-rank graph matching context, we may obtain estimated latent node positions by directly decomposing the adjacency matrices:
\begin{align}
A^{(1)} &\approx \hat{X}^{(1)} \hat{J}^{(1)} \left(\hat{X}^{(1)}\right)^T   \label{A1::decompose}\\
A^{(2)} &\approx \hat{X}^{(2)} \hat{J}^{(2)} \left(\hat{X}^{(2)}\right)^T  \label{A2::decompose}
\end{align}
where $\hat{X}^{(\ell)}\in\mathbb{R}^{n\times d}$ and $\hat{J}^{(\ell)}\in \textrm{diag}(\{\pm1\}^d)$.  We then solve the following point registration problem:
\begin{equation}
\min_{P,O}\|P\hat{X}^{(1)}O-\hat{X}^{(2)}\|_F
\label{problem::reduced::graph_matching}
\end{equation}
for permutation matrix $P\in\{0,1\}^{n\times n}$ and $O\in{\cal O}^d\subset \mathbb{R}^{d\times d}$. 

Computationally, our method demands optimization of the function $\widehat{\Delta}(O;X,Y;R)$ over $O\in{\cal O}^d$, where ${\cal O}^d$ is the collection of all $d\times d$ orthonormal matrices.  For each $O$, the cost to evaluate $\widehat{\Delta}$ is $O((n_X+n_Y)m_s)$, where recall that $n_X$ and $n_Y$ are sample sizes of the data sets and we have control over $m_s$, the number of $s$'es we shall sample element-wise from $\pi(t)$.  This contrasts the $O(n^3)$ cost of estimating the best match within each iteration in ICP, and moreover gives us the flexibility of controlling the trade-off between computation time and accuracy.  Compared to KC, our method can handle continuous, discrete or mixed distributions by a unified formulation.  Compared to both ICP and KC, our method is backed by a consistency guarantee with an explicit error rate.  The results will be presented in Section \ref{section::theory}.

Before concluding the description of our method, we briefly explain two small but important details.  First, our criterion \eqref{definition::loss_func_sample} does not require equal sample sizes.  If the sample sizes are different, we may simply bootstrap the smaller sample, and the Hungarian algorithm will naturally produce a many-to-one estimated map, which is desired.  The second topic is the choice of $R$.  At first it may seem natural to choose $R$ as an increasing function of $n$, as did in an earlier version of this paper.  However, doing so would likely greatly depreciate the guaranteed error rate.  If we recall the idea of matching moments that motivated our method, we realize an arguable intuition that all the moments can be determined by the curvature of the Laplace transform around $\im(s)\approx 0$, and as we travel far away with large $\|\im(s)\|_2$ in the tail, the shape of the Laplace transforms there might possibly grow less relevant.  In Section \ref{section::theory}, we shall see that fixing $R$ helps us to achieve a nearly tight error bound.

\section{Theory}
\label{section::theory}

By \citet{zhang2014detecting} and \citet{anderson1986introduction}, with probability $1-\delta(n)$ where $\delta(x)\to0$ as $x\to\infty$, for the $k$th largest nonzero eigenvalue of the matrix $W$, denoted by $\lambda_k(W)$, we have
\begin{align}
\frac{\lambda_k(W)}{n\lambda_k(f)} &\asymp C_k>0  \label{eigen::concentration::1}\\
\left|\hat{\lambda}_k(W)-\lambda_k(W)\right| &\preceq \tilde{C}_k\sqrt{n}  \label{eigen::concentration::2}
\end{align}
Without loss of generality, we may organize the columns of $X$ and $\hat{X}^{(\ell)}$ in \eqref{A1::generate}, \eqref{A2::generate} and \eqref{A1::decompose}, \eqref{A2::decompose} to be put in the order aligned to the true or estimated leading nonzero eigenvalues of the corresponding matrices from which those $X$'s are decomposed, then by \eqref{eigen::concentration::1} and \eqref{eigen::concentration::2}, we have
$$
\pr\left( \{ \hat{J}^{(1)}\neq J \}\cup \{ \hat{J}^{(2)}\neq J \} \right) \leq 2\delta(n)\to0
$$
Next, combining the results of \citet{yu2014useful}, \citet{lei2015consistency}, \eqref{eigen::concentration::1} and \eqref{eigen::concentration::2}, there exists unknown orthonormal matrices $\hat{O}_X, \hat{O}_Y\in{\cal O}^d$, such that with high probability,
\begin{align}
\|\hat{X}^{(1)}\hat{O}_X-X\|_F &\leq C_1  \label{X1::Davis-Kahan}\\
\|\hat{X}^{(2)}\hat{O}_Y-Y\|_F &\leq C_2  \label{X2::Davis-Kahan}
\end{align}
Moreover, if $J=$~BlockDiagonal$(I_{d_1},-I_{d_2})$, then we may further shrink the sample spaces of $\hat{O}_X$ and $\hat{O}_Y$ as follows:
$$
\hat{O}_X = \begin{pmatrix}
\hat{O}_X^+ &0\\0&\hat{O}_X^-
\end{pmatrix}\quad\textrm{and}\quad
\hat{O}_Y = \begin{pmatrix}
\hat{O}_Y^+ &0\\0&\hat{O}_Y^-
\end{pmatrix}
$$
where  $\hat{O}_X^+, \hat{O}_Y^+\in{\cal O}^{d_1}$, and $\hat{O}_X^-, \hat{O}_Y^-\in {\cal O}^{d_2}$, respectively, because we can apply \citet{yu2014useful} on positive and negative eigenvalues and their corresponding eigenvectors, respectively.  This reduction was not explicitly emphasized in network analysis literature, mostly works on community detection, because the orthonormal transform $\hat{O}$ is nuisance and has no impact on the subsequent estimation and inference steps.  But in the matching problem, the dimensionality of $O$ is determining on both accuracy and computation cost.

We now present the consistency theory of our method.  The proofs are in the Appendix.  First we present the uniform concentration inequality of our proposed criterion to its population version.

\begin{theorem}[Uniform concentration of the loss function]
	\label{theorem::theorem_1}
	Given the distributions of universally bounded random vectors $\X$ and $\Y$ in $\mathbb{R}^d$, there are universal constants $C,r_s,r_X,r_Y>0$ such that
	\begin{align}
	&\pr\Bigg( \sup_{O\in{\cal O}^d} \left| \widehat{\Delta}(O;X,Y;R) - \Delta(O;\X,\Y;R) \right|  \notag\\
	& >C\left\{ \widetilde{C}(R;\gamma) \right\}^d\left\{ \sqrt{\frac{\log R+\log m_s}{m_s}} + \sqrt{\frac{\log R+\log n_X}{n_X}} + \sqrt{\frac{\log R+\log n_Y}{n_Y}} \right\}  \Bigg)\notag\\
	&\leq C\left\{ (Rm_S)^{-r_S}+(Rn_X)^{-r_X}+(Rn_Y)^{-r_Y} \right\}
	\label{eqn::theorem_1}
	\end{align}
	Moreover,
	\begin{align}
	&\pr\Bigg( \sup_{O\in{\cal O}^d} \left| \widehat{\Delta}(O;\widehat{X}\widehat{O}_X,\widehat{Y}\widehat{O}_Y;R) - \Delta(O;\X,\Y;R) \right|  \notag\\
	& >C\left\{ \widetilde{C}(R;\gamma) \right\}^d\left\{ \sqrt{\frac{\log R+\log m_s}{m_s}} + \sqrt{\frac{\log R+\log n_X}{n_X}} + \sqrt{\frac{\log R+\log n_Y}{n_Y}} \right\}   + C(R)\left(\frac1{\sqrt{n_X}}+\frac1{\sqrt{n_Y}}\right)    \Bigg)\notag\\
	&\leq C\left\{ (Rm_S)^{-r_S}+(Rn_X)^{-r_X}+(Rn_Y)^{-r_Y} \right\}
	\label{eqn::theorem_1_measurement_error}
	\end{align}
	where $C(R)$ is a constant depending only on $R$.
\end{theorem}

It is worth noting that \eqref{eqn::theorem_1_measurement_error} is stated with unknown transforms $\widehat{O}_X$ and $\widehat{O}_Y$.  This means that we know that with high probability $\widehat{\Delta}(\widehat{O}_XO\widehat{O}_Y^T;\widehat{X}, \widehat{Y}; R)$ and $\Delta(O;\X,\Y;R)$ will be close, but we cannot disentangle $\widehat{O}_X$ and $\widehat{O}_Y$ mixed inside the optimal $\widehat{O}:=\arg\min_O\widehat{\Delta}(O;\widehat{X},\widehat{Y};R)$, but this is fine.  Recall that our ultimate goal is to accurately estimate the point registration $\hat{P}$.  Our theory only demands that $\widehat{O}_X^T\widehat{O}\widehat{O}_Y$ is close to some optimal solution $O^*$, if the optimal solution is not unique.

Next we state a crucial regularity condition regarding the shape of our loss function near its minimum.  This property is satisfied by a wide range of frequently used distributions in practice.

\begin{definition}[Sharp Slope Condition]
	\label{definition::SSC}
	A function $\Gamma(\alpha_1,\ldots,\alpha_d)$ is said to satisfy \emph{Steep Slope Condition}, if there exist universal constants $\delta_0>0, C_1>0, C_2>0$ such that for all $\alpha=(\alpha_1,\ldots,\alpha_d)\in{\cal A}$, for some compact ${\cal A}\subset \mathbb{R}^d$, the following properties hold:
	\begin{itemize}
		\item The function $\Gamma$ is minimized to 0:$$\min_\alpha \Gamma(\alpha)=0$$ and the minimum is attained only at a finite number of $\alpha$'s
		\item For any $$\alpha^* \in \arg\min_\alpha \Gamma(\alpha)$$ and $\alpha\in B(\alpha^*;\delta_0)$, we have
		$$
		C_1\|\alpha-\alpha^*\|_2\leq \Gamma(\alpha) \leq C_2\|\alpha-\alpha^*\|_2
		$$
	\end{itemize}
\end{definition}

The Sharp Slope Condition is satisfied by $\Delta(O;\X,\Y;R)$ with respect to $O$ restricted in orthonormal transformations for many distributions, examples include multinomial distribution and multivariate normal distribution -- notice the latter is not within the range of the consideration of our current theory as the distribution is unbounded, but we believe the it can be expanded to sub-gaussian distributions. 
If a function satisfies Sharp Slope Condition, then optimizing a sample version of the function that has uniform convergence would yield an estimation decently close to the true optimal solution.

\begin{theorem}
	\label{theorem::theorem_2}
	Suppose the function $\Gamma(\alpha_1,\ldots,\alpha_d)$ satisfies Sharp Slope Condition, and it has a uniformly concentrating sample version $\widehat{\Gamma}(\alpha)$ such that
	$$
	\sup_{\alpha\in{\cal A}}\left| \widehat{\Gamma}(\alpha)-\Gamma(\alpha) \right|\leq \varepsilon
	$$
	Assume that $\log(1/\varepsilon)\asymp \log n$ and the time cost to evaluate $\widehat{\Gamma}(\alpha)$ is polynomial in $n$, when $n$ is the sample size associated with $\widehat{\Gamma}$.  Then there exists a polynomial algorithm, such its output
	$$
	\hat{\alpha} = \widehat{\arg}\min_{\alpha}|\widehat{\Gamma}(\alpha)|
	$$
	is close to the optimal solution, in the sense that
	$$
	\min_{\alpha^*:\Gamma(\alpha^*)=0}\|\hat{\alpha}-\alpha^*\|_2\leq C\varepsilon
	$$
\end{theorem}

If $\Delta(O;\X,\Y;R)$ satisfies Sharp Slope Condition, where we can regard $\Gamma=\Delta$ and $\alpha$ to be some parameterization of $O$, such as when $d=2$ we can parameterize $O\in$~SO$(2)$ by the rotation angle $\theta$, then using the results of \citet{levina2001earth}, \citet{fournier2015rate} and \citet{lei2018convergence}, for the equal sample size case $n_X=n_Y=n$, we have

\begin{theorem}
	\label{theorem::theorem_3}
	Suppose the parameterization of $\Delta(O;\X,\Y;R)$ as a function of $O$ satisfies the Sharp Slope Condition.  Let $\widehat{O}$ be the output of the algorithm in Theorem \ref{theorem::theorem_2}, and then define $\widehat{P}$ to be the optimal permutation under MSE estimated by the Hungarian algorithm to match the rows of $X\widehat{O}$ and $Y$, we have
	$$
	\min_O\frac{\|\widehat{P}XO-Y\|_F}{\sqrt{n}} = O_p\left( n^{-1/\max(d,2)} \right)
	$$
\end{theorem}

Notice that when $d\geq 2$, the term $n^{-1/d}$ is dominating.  The error bound by Theorem \ref{theorem::theorem_3} seems tight when $d\geq 2$ among methods that assume population structures, as the concentration of the empirical distribution to the population distribution is likely unavoidable.

Theorem \ref{theorem::theorem_3} immediately implies the control on graph matching error:
\begin{equation}
	\frac1n \|\widehat{P}W^{(1)}\widehat{P}-W^{(2)}\|_F  = O_p\left( n^{-1/\max(d, 2)} \right)
\end{equation}

\section{Numerical examples}

In this section, we test our method and two other popular benchmark methods for unseeded graph matching on three example low-rank graphs.  Graph 1 is generated from the graphon of a stochastic block model with $K=4$ communities of equal sizes.  Within-community probabilities of community $i$ is $i/5$ and between-community probabilities are $0.3/5$.  Graphon 2 is a more general low-rank graph with distinct nonzero eigenvalues.  The graphon function is defined by $f(x,y) = \sin(5\pi(x+y+1))/2 + 0.5$.  Graphon 3 is relatively most difficult for all methods, as it has repeated nonzero eigenvalues.  The leading eigenvalues are $(0.167,0.05,0.05,0.05)$ and their corresponding eigenfunctions are $\xi_1(x)=1$, $\xi_2(x)=2x-1$, $\xi_3(x)=1-4|x-0.5|$ and $\xi_4(x)=2\cdot\mathbbm{1}_{[(1/8,3/8)\cup(5/8,7/8)]}-1$.  Graphs generated from graphon 1 element-wise follow Bernoulli distributions, and graphs generated from both graphons 2 and 3 are element-wise contaminated by $N(0,0.2)$ random noises.  We repeat the experiment 30 times for each graphon.  In each experiment, we randomly generate two independent realizations from the graphon, and shuffle the node order of one of the adjacency matrices by a randomly chosen permutation matrix $P^*$ unknown to all the compared methods.  We measure the performance of the methods by RMSE:
$$
\textrm{RMSE} = \|\hat{P}W\hat{P}^T - P^*W(P^*)^T\|_F/n
$$
In all these experiments, we run our method with the following random starts:  we initialize the orthonormal matrix $O$ in our loss function $\hat{\Delta}(O;X,Y;R)$ from Givens rotations \citep{merchant2018efficient}:
$$
{\cal G}(u;\theta_1,\ldots,\theta_1{d-1}):= G_{d-1}(\theta_{d-1})\cdots G_1(\theta_1)G_0(u)
$$
Where $G_0\cdots G_{d-1}\in\mathbb{R}^{d\times d}$ are defined as follows; $G_0=\textrm{diag}(u,1,\ldots,1)$ where $u=\pm1$ and $(G_k)_{1,1}(\theta_k)=(G_k)_{k+1,k+1}(\theta_k)=\cos(\theta_k)$, $(G_k)_{k+1,1}(\theta_k)=-(G_k)_{1,k+1}(\theta_k)=\sin(\theta_k)$ and the $G_k$ matrix excluding the $1$st and $k$th rows and columns is an identity matrix $(G_k)_{(-1,-(k+1)),(-1,-(k+1))}=I_{d-2}$.  We start our method with $(u,\theta_1,\ldots,\theta_{d-1})\in \{\pm1\}\times \{ 0,1/p,\ldots,(p-1)/p \}^{d-1}$, where we choose $p=4$.  Notice that this is feasible since all the tested graphons have at most $d=4$ nonzero leading eigenvalues.  If $d$ is large, we may reduce $p$ to $p=2,3$ and also considering sub-sampling from all the possible configurations of $u$ and $\theta$'s.  In this simulation study, we fix $m_s=500$ and $R=15$.  For $n=2000$, we used LAPJV to perform the final Hungarian algorithm match, and for all the other (smaller) $n$'s, we used MUNKRES.

The bench mark methods we compared to are \citet{fishkind2012seeded} and \citet{vogelstein2015fast} using the MATLAB codes downloaded from the authors' websites.

\begin{table}[h!]
	\centering
	\caption{Graphon 1 (100*Frobenius/n, repeat = 30 times)}
	\begin{tabular}{ccccc}\hline
		Graphon & net size & Our method & SGM & FAQ\\\hline
		\multirow{5}{*}{RMSE}	& $n=100$ & 5.02(0.28)	&6.03(0.24)	 &31.48(0.01)	\\
		& $n=250$ & 0.00(0.09)	&0.00(0.05)	 &31.89(0.00)\\
		& $n=500$ & 0.00(0.00)	&0.00(0.00)	 &31.59(0.00)\\
		& $n=1000$ & 0.00(0.00)	 & 0.00(0.00)	& 31.60(0.00)\\
		& $n=2000$ & 0.00(0.00)	 & 0.00(0.00)	& 31.61(0.00)\\
		\hline
		\multirow{5}{*}{Time}		& $n=100$ & 5.57(0.36)&  0.14(0.00)&  0.17(0.00)\\
		& $n=250$ & 19.57(0.07)&  0.96(0.01)&  1.69(0.01)\\
		& $n=500$ &27.25(0.06)&  4.87(0.02)&  8.64(0.02)\\
		& $n=1000$ & 34.84(0.08)& 25.22(0.04)& 49.98(0.08)\\
		& $n=2000$ & 55.31(0.08)&141.21(0.11)&327.48(0.26)\\\hline
	\end{tabular}
\end{table}

\begin{table}[h!]
	\centering
	\caption{Graphon 2 (100*Frobenius/n, repeat = 30 times)}
	\begin{tabular}{ccccc}\hline
		Graphon & net size & Our method & SGM & FAQ\\\hline
		\multirow{5}{*}{RMSE}	& $n=100$ & 8.62(0.17)&  6.99(0.06)& 70.30(0.01)\\
		& $n=250$ &6.18(0.11)&  4.96(0.02)& 70.52(0.00)\\
		& $n=500$ & 4.38(0.04)&  3.91(0.01)& 70.60(0.00)\\
		& $n=1000$ & 3.06(0.02)&  3.18(0.00)& 70.64(0.00)\\
		& $n=2000$ & 2.35(0.01)&  2.46(0.00)& 70.67(0.00)\\
		\hline
		\multirow{5}{*}{Time}		& $n=100$ & 5.33(0.02)&  0.16(0.00)&  0.28(0.00)\\
		& $n=250$ & 12.55(0.03)&  1.16(0.00)&  1.90(0.01)\\
		& $n=500$ &18.20(0.10)&  5.96(0.01)&  9.72(0.02)\\
		& $n=1000$ & 29.26(0.07)& 30.67(0.03)& 57.29(0.04)\\
		& $n=2000$ &50.30(0.04)&189.78(0.14)&376.96(0.18)\\\hline
	\end{tabular}
\end{table}

\begin{table}[h!]
	\centering
	\caption{Graphon 3 (100*Frobenius/n, repeat = 30 times)}
	\begin{tabular}{ccccc}\hline
		Graphon & net size & Our method & SGM & FAQ\\\hline
		\multirow{5}{*}{RMSE}	& $n=100$ & 3.10(0.05)&  5.88(0.20)& 11.63(0.06)\\
		& $n=250$ & 2.46(0.07)&  5.74(0.17)& 11.62(0.02)\\
		& $n=500$ & 1.99(0.10)&  6.03(0.15)& 11.64(0.01)\\
		& $n=1000$ & 1.74(0.08)& 12.23(0.13)& 11.88(0.01)\\
		& $n=2000$ & 1.51(0.07)& 12.24(0.03)& 11.99(0.00)\\
		\hline
		\multirow{5}{*}{Time}		& $n=100$ & 18.81(0.38)&  0.14(0.00)&  0.82(0.01)\\
		& $n=250$ & 42.30(0.45)&  1.01(0.02)&  5.74(0.03)\\
		& $n=500$ &68.00(0.56)&  5.95(0.09)& 27.87(0.17)\\
		& $n=1000$ & 109.94(0.61)& 16.95(0.32)&159.75(0.62)\\
		& $n=2000$ & 194.93(0.87)&121.93(0.39)&1028.84(2.98)\\\hline
	\end{tabular}
\end{table}

Graphon 1 is relatively easy for all methods, and we observed that our method and SGM quickly became perfectly accurate from a quite small sample size $(n=250)$ onward.  Graphon 2 is slightly harder and our method performed similarly to SGM.  On the most challenging model Graphon 3, our method shows its advantage in exploiting the low-rank structure and has a diminishing MSE, whereas SGM seemed to become increasingly disoriented in its growing search space of optimization.  In all examples, FAQ did not perform well, possibly due to the poor initialization at $1/n\cdot \mathbbm{1}\mathbbm{1}^T$.

On computational efficiency, we observe that with a fixed $m_s$, our method's time complexity increases linearly with the sample size $n$.  Recall that we have the flexibility to tune the increment rate of $m_s$ with $n$, so in the extreme case if we have to handle an increasing sample size with a fixed target error bound, then we may use a fixed $m_s$ to achieve linear computational time.  Also recall that increasing $m_s$ faster than $n$, however, will not further improve error rate since $n$ is now the bottleneck factor.

\section{Discussions}

In this section, we present discussions on various aspects of our method and some future directions along the two lines of point registration and graph matching.

First, on the point registration side:  our method can handle more general invertible linear transformations than orthonormal $O$, but in order to retain the satisfaction of the Steep Slope Condition and our analysis, some regularity assumptions on the family of transformations would be necessary.  Similarly, further extension to parameterized nonlinear transformation can be considered.  We envision the even further extension to general nonparametric transformations to remain challenging.

In this paper, we have been placing our attentions solely to matching points from universally bounded distributions, which is indeed a natural feature of positions by decomposing moderately regular graphons.  We conjecture that our theoretical results might be generalizable to sub-Gaussian $\X$ and $\Y$ and possibly other light-tailed distributions, as many preliminary empirical evidences encouragingly suggest, and we are currently working on this direction.  On the other hand, if the distributions to be matched are extremely heavy-tailed that even the first moment does not exist, then we may need a new goodness measurement as the population Wasserstein distance may not be always well-defined unless the two distributions are already perfectly matched up.  If the population version of some criterion is not-defined, the meaningfulness of its sample version, despite its possible existence, would be under doubts.

Matching the points generated from different graphons, however, remains a major challenge.  Notice that our criterion as a discrepancy measure between the two distributions' Laplace transforms is, by itself, a valid statistical distance, as it satisfies triangular inequality and other requirements for a distance.  With non-matchable distributions, optimizing our criterion and optimizing other criteria such as Wasserstein distance may find different estimated transforms of one data set that ``best matches'' the other in their own senses, and subsequently, the resulting matches are likely different.  The potential presence of outliers is a similar but different topic -- the two underlying point generating distributions may still be matchable, but now except for a few outliers.  Another closely related but different topic is outliers -- our method might not be robust against outliers, and we recommend users to detect and eliminate outliers in the data pre-processing procedure.

Then, on the graph matching side.  First, we made the assumption that there are optimal $P^*$ and $O^*$ that can perfectly match up the true latent node positions, that is, $P^*XO^*=Y$, under equal sample sizes.  This assumption is by no means substantive and the assumption could be easily relaxed to networks whose latent node positions in the graphon model $u_1,\ldots,u_{n_X}$ and $v_1,\ldots,v_{n_Y}$ or arbitrary dependence structure between each pair of nodes belonging to different networks, while the internal independence within each network holds: $u_i\perp u_j$ and $v_{i'}\perp v_{j'}$ for any $i\neq j$ and $i'\neq j'$.

Our method is certainly not designed to match up general graphons of full rank, despite it may find competitive solutions under some full rank graphons, if the leading few eigenfunctions can already differentiate the different roles of the nodes in the networks well.  In such cases, despite the other eigenvalues and eigenfunctions may matter much for purposes such as graphon estimation, the leading ones may already suffice to provide accurate node matching.  But the general problem of matching full rank graphons is outside the scope of this paper.


\bibliographystyle{abbrvnat}
\bibliography{main}

\begin{thebibliography}{36}
\providecommand{\natexlab}[1]{#1}
\providecommand{\url}[1]{\texttt{#1}}
\expandafter\ifx\csname urlstyle\endcsname\relax
  \providecommand{\doi}[1]{doi: #1}\else
  \providecommand{\doi}{doi: \begingroup \urlstyle{rm}\Url}\fi

\bibitem[Airoldi et~al.(2013)Airoldi, Costa, and Chan]{airoldi2013stochastic}
E.~M. Airoldi, T.~B. Costa, and S.~H. Chan.
\newblock Stochastic blockmodel approximation of a graphon: Theory and
  consistent estimation.
\newblock In \emph{Advances in Neural Information Processing Systems}, pages
  692--700, 2013.

\bibitem[Aldous(1981)]{aldous1981representations}
D.~J. Aldous.
\newblock Representations for partially exchangeable arrays of random
  variables.
\newblock \emph{Journal of Multivariate Analysis}, 11\penalty0 (4):\penalty0
  581--598, 1981.

\bibitem[Anderson et~al.(1986)Anderson, Aulin, and
  Sundberg]{anderson1986introduction}
J.~B. Anderson, T.~Aulin, and C.-E. Sundberg.
\newblock Introduction.
\newblock In \emph{Digital Phase Modulation}, pages 1--14. Springer, 1986.

\bibitem[Athreya et~al.(2017)Athreya, Fishkind, Levin, Lyzinski, Park, Qin,
  Sussman, Tang, Vogelstein, and Priebe]{athreya2017statistical}
A.~Athreya, D.~E. Fishkind, K.~Levin, V.~Lyzinski, Y.~Park, Y.~Qin, D.~L.
  Sussman, M.~Tang, J.~T. Vogelstein, and C.~E. Priebe.
\newblock Statistical inference on random dot product graphs: a survey.
\newblock \emph{arXiv preprint arXiv:1709.05454}, 2017.

\bibitem[Babai(2016)]{babai2016graph}
L.~Babai.
\newblock Graph isomorphism in quasipolynomial time.
\newblock In \emph{Proceedings of the forty-eighth annual ACM symposium on
  Theory of Computing}, pages 684--697. ACM, 2016.

\bibitem[Bickel and Chen(2009)]{bickel2009nonparametric}
P.~J. Bickel and A.~Chen.
\newblock A nonparametric view of network models and newman--girvan and other
  modularities.
\newblock \emph{Proceedings of the National Academy of Sciences}, pages
  pnas--0907096106, 2009.

\bibitem[Conte et~al.(2004)Conte, Foggia, Sansone, and Vento]{conte2004thirty}
D.~Conte, P.~Foggia, C.~Sansone, and M.~Vento.
\newblock Thirty years of graph matching in pattern recognition.
\newblock \emph{International journal of pattern recognition and artificial
  intelligence}, 18\penalty0 (03):\penalty0 265--298, 2004.

\bibitem[Du et~al.(2010)Du, Zheng, Ying, and Liu]{du2010affine}
S.~Du, N.~Zheng, S.~Ying, and J.~Liu.
\newblock Affine iterative closest point algorithm for point set registration.
\newblock \emph{Pattern Recognition Letters}, 31\penalty0 (9):\penalty0
  791--799, 2010.

\bibitem[Durrett(2010)]{durrett2010probability}
R.~Durrett.
\newblock \emph{Probability: theory and examples}.
\newblock Cambridge university press, 2010.

\bibitem[Ezra et~al.(2006)Ezra, Sharir, and Efrat]{ezra2006icp}
E.~Ezra, M.~Sharir, and A.~Efrat.
\newblock On the icp algorithm.
\newblock In \emph{Proceedings of the twenty-second annual symposium on
  Computational geometry}, pages 95--104. ACM, 2006.

\bibitem[Fishkind et~al.(2012)Fishkind, Adali, Patsolic, Meng, Lyzinski, and
  Priebe]{fishkind2012seeded}
D.~E. Fishkind, S.~Adali, H.~G. Patsolic, L.~Meng, V.~Lyzinski, and C.~E.
  Priebe.
\newblock Seeded graph matching.
\newblock \emph{arXiv preprint arXiv:1209.0367}, 2012.

\bibitem[Fournier and Guillin(2015)]{fournier2015rate}
N.~Fournier and A.~Guillin.
\newblock On the rate of convergence in wasserstein distance of the empirical
  measure.
\newblock \emph{Probability Theory and Related Fields}, 162\penalty0
  (3-4):\penalty0 707--738, 2015.

\bibitem[Gao et~al.(2015)Gao, Lu, Zhou, et~al.]{gao2015rate}
C.~Gao, Y.~Lu, H.~H. Zhou, et~al.
\newblock Rate-optimal graphon estimation.
\newblock \emph{The Annals of Statistics}, 43\penalty0 (6):\penalty0
  2624--2652, 2015.

\bibitem[Hoover(1979)]{hoover1979relations}
D.~N. Hoover.
\newblock Relations on probability spaces and arrays of random variables.
\newblock \emph{Preprint, Institute for Advanced Study, Princeton, NJ}, 2,
  1979.

\bibitem[Lei(2018{\natexlab{a}})]{lei2018convergence}
J.~Lei.
\newblock Convergence and concentration of empirical measures under wasserstein
  distance in unbounded functional spaces.
\newblock \emph{arXiv preprint arXiv:1804.10556}, 2018{\natexlab{a}}.

\bibitem[Lei(2018{\natexlab{b}})]{lei2018network}
J.~Lei.
\newblock Network representation using graph root distributions.
\newblock \emph{arXiv preprint arXiv:1802.09684}, 2018{\natexlab{b}}.

\bibitem[Lei et~al.(2015)Lei, Rinaldo, et~al.]{lei2015consistency}
J.~Lei, A.~Rinaldo, et~al.
\newblock Consistency of spectral clustering in stochastic block models.
\newblock \emph{The Annals of Statistics}, 43\penalty0 (1):\penalty0 215--237,
  2015.

\bibitem[Levina and Bickel(2001)]{levina2001earth}
E.~Levina and P.~Bickel.
\newblock The earth mover's distance is the mallows distance: Some insights
  from statistics.
\newblock In \emph{null}, page 251. IEEE, 2001.

\bibitem[Lyzinski(2016)]{lyzinski2016information}
V.~Lyzinski.
\newblock Information recovery in shuffled graphs via graph matching.
\newblock \emph{arXiv preprint arXiv:1605.02315}, 2016.

\bibitem[Lyzinski et~al.()Lyzinski, Fishkind, Fiori, Vogelstein, Priebe, and
  Sapiro]{lyzinski2016graph}
V.~Lyzinski, D.~Fishkind, M.~Fiori, J.~Vogelstein, C.~Priebe, and G.~Sapiro.
\newblock Graph matching: Relax at your own risk.
\newblock \emph{IEEE Transactions on Pattern Analysis \& Machine Intelligence},
  \penalty0 (1):\penalty0 1--1.

\bibitem[Lyzinski et~al.(2014{\natexlab{a}})Lyzinski, Adali, Vogelstein, Park,
  and Priebe]{lyzinski2014seeded2}
V.~Lyzinski, S.~Adali, J.~T. Vogelstein, Y.~Park, and C.~E. Priebe.
\newblock Seeded graph matching via joint optimization of fidelity and
  commensurability.
\newblock \emph{arXiv preprint arXiv:1401.3813}, 2014{\natexlab{a}}.

\bibitem[Lyzinski et~al.(2014{\natexlab{b}})Lyzinski, Fishkind, and
  Priebe]{lyzinski2014seeded}
V.~Lyzinski, D.~E. Fishkind, and C.~E. Priebe.
\newblock Seeded graph matching for correlated erd{\"o}s-r{\'e}nyi graphs.
\newblock \emph{Journal of Machine Learning Research}, 15\penalty0
  (1):\penalty0 3513--3540, 2014{\natexlab{b}}.

\bibitem[Lyzinski et~al.(2014{\natexlab{c}})Lyzinski, Sussman, Fishkind, Pao,
  Vogelstein, and Priebe]{lyzinski2014seeded3}
V.~Lyzinski, D.~L. Sussman, D.~E. Fishkind, H.~Pao, J.~T. Vogelstein, and C.~E.
  Priebe.
\newblock Seeded graph matching for large stochastic block model graphs.
\newblock \emph{stat}, 1050:\penalty0 12, 2014{\natexlab{c}}.

\bibitem[Maron et~al.(2016)Maron, Dym, Kezurer, Kovalsky, and
  Lipman]{maron2016point}
H.~Maron, N.~Dym, I.~Kezurer, S.~Kovalsky, and Y.~Lipman.
\newblock Point registration via efficient convex relaxation.
\newblock \emph{ACM Transactions on Graphics (TOG)}, 35\penalty0 (4):\penalty0
  73, 2016.

\bibitem[Merchant et~al.(2018)Merchant, Vatwani, Chattopadhyay, Raha, Nandy,
  Narayan, and Leupers]{merchant2018efficient}
F.~Merchant, T.~Vatwani, A.~Chattopadhyay, S.~Raha, S.~Nandy, R.~Narayan, and
  R.~Leupers.
\newblock Efficient realization of givens rotation through
  algorithm-architecture co-design for acceleration of qr factorization.
\newblock \emph{arXiv preprint arXiv:1803.05320}, 2018.

\bibitem[Post(1930)]{post1930generalized}
E.~L. Post.
\newblock Generalized differentiation.
\newblock \emph{Transactions of the American Mathematical Society}, 32\penalty0
  (4):\penalty0 723--781, 1930.

\bibitem[Svensson and Tarnawski(2017)]{svensson2017matching}
O.~Svensson and J.~Tarnawski.
\newblock The matching problem in general graphs is in quasi-nc.
\newblock In \emph{Foundations of Computer Science (FOCS), 2017 IEEE 58th
  Annual Symposium on}, pages 696--707. Ieee, 2017.

\bibitem[Udell and Townsend(2017)]{udell2017nice}
M.~Udell and A.~Townsend.
\newblock Nice latent variable models have log-rank.
\newblock \emph{arXiv preprint arXiv:1705.07474}, 2017.

\bibitem[Vogelstein et~al.(2015)Vogelstein, Conroy, Lyzinski, Podrazik,
  Kratzer, Harley, Fishkind, Vogelstein, and Priebe]{vogelstein2015fast}
J.~T. Vogelstein, J.~M. Conroy, V.~Lyzinski, L.~J. Podrazik, S.~G. Kratzer,
  E.~T. Harley, D.~E. Fishkind, R.~J. Vogelstein, and C.~E. Priebe.
\newblock Fast approximate quadratic programming for graph matching.
\newblock \emph{PLOS one}, 10\penalty0 (4):\penalty0 e0121002, 2015.

\bibitem[Widder(2015)]{widder2015laplace}
D.~V. Widder.
\newblock \emph{Laplace transform (PMS-6)}.
\newblock Princeton university press, 2015.

\bibitem[Wolfe and Olhede(2013)]{wolfe2013nonparametric}
P.~J. Wolfe and S.~C. Olhede.
\newblock Nonparametric graphon estimation.
\newblock \emph{arXiv preprint arXiv:1309.5936}, 2013.

\bibitem[Xu(2017)]{xu2017rates}
J.~Xu.
\newblock Rates of convergence of spectral methods for graphon estimation.
\newblock \emph{arXiv preprint arXiv:1709.03183}, 2017.

\bibitem[Young and Scheinerman(2007)]{young2007random}
S.~J. Young and E.~R. Scheinerman.
\newblock Random dot product graph models for social networks.
\newblock In \emph{International Workshop on Algorithms and Models for the
  Web-Graph}, pages 138--149. Springer, 2007.

\bibitem[Yu et~al.(2014)Yu, Wang, and Samworth]{yu2014useful}
Y.~Yu, T.~Wang, and R.~J. Samworth.
\newblock A useful variant of the davis--kahan theorem for statisticians.
\newblock \emph{Biometrika}, 102\penalty0 (2):\penalty0 315--323, 2014.

\bibitem[Zhang et~al.(2014)Zhang, Levina, and Zhu]{zhang2014detecting}
Y.~Zhang, E.~Levina, and J.~Zhu.
\newblock Detecting overlapping communities in networks using spectral methods.
\newblock \emph{arXiv preprint arXiv:1412.3432}, 2014.

\bibitem[Zhang et~al.(2017)Zhang, Levina, and Zhu]{zhang2017estimating}
Y.~Zhang, E.~Levina, and J.~Zhu.
\newblock Estimating network edge probabilities by neighbourhood smoothing.
\newblock \emph{Biometrika}, 104\penalty0 (4):\penalty0 771--783, 2017.

\end{thebibliography}

\end{document}